\documentclass[prd,preprint,superscriptaddress,showpacs,byrevtex]{revtex4}
\usepackage{bm}
\def\fsl#1{\setbox0=\hbox{$#1$}           
   \dimen0=\wd0                                 
   \setbox1=\hbox{/} \dimen1=\wd1               
   \ifdim\dimen0>\dimen1                        
      \rlap{\hbox to \dimen0{\hfil/\hfil}}      
      #1                                        
   \else                                        
      \rlap{\hbox to \dimen1{\hfil$#1$\hfil}}   
      /                                         
   \fi}                                         %
\newcommand{\be}{\begin{equation}}
\newcommand{\ee}{\end{equation}}
\newcommand{\bea}{\begin{eqnarray}}
\newcommand{\eea}{\end{eqnarray}}
\newcommand{\beq}{\begin{equation}}
\newcommand{\eeq}{\end{equation}}
\newcommand{\beqs}{\begin{eqnarray}}
\newcommand{\eeqs}{\end{eqnarray}}

\newcommand{\aslash}{A\hspace{-0.067in}\slash}

\begin{document}
\title{ Lattice QCD Method To Study Parton To Hadron Fragmentation Function }
\author{Gouranga C Nayak }\thanks{E-Mail: nayakg138@gmail.com}
%
%
\date{\today}
\begin{abstract}
In the literature it is assumed that the parton to hadron fragmentation function cannot be studied by using the lattice QCD method because of the sum over the (unobserved) outgoing hadronic states. However, in this paper we find that since the hadron formation from the partons can be studied by using the lattice QCD method, the parton to hadron fragmentation function can be studied by using the lattice QCD method by using the LSZ reduction formula for the partonic processes.
\end{abstract}
\pacs{13.87.Fh, 12.38.Gc, 14.40.-n, 14.20.-c }
\maketitle
\pagestyle{plain}

\pagenumbering{arabic}

\section{ Introduction }

At high energy colliders two incoming hadrons (or leptons or lepton-hadron) collide to produce a bunch of particles which can be hadrons or non-hadrons (such as leptons, photons etc.). At the large momentum transfer the interaction involving the quarks and gluons processes occur at the high energy colliders. Since we have not directly experimentally observed quarks and gluons these quarks and gluons fragment to hadrons which are directly experimentally observed. Hence the parton to hadron fragmentation function plays an important role at the high energy colliders. The parton to hadron fragmentation function also plays an important role to detect the quark-gluon plasma at RHIC and LHC \cite{qg1,qg2}.

The interaction between quarks and gluons is described by the quantum chromodynamics (QCD) \cite{ymw} which is a fundamental theory of the nature. Due to the asymptotic freedom in QCD \cite{gww} the renormalized QCD coupling decreases at small distance where the perturbative QCD (pQCD) is applicable. Hence the short distance partonic scattering cross section at the high energy colliders can be calculated by using the pQCD. This short distance partonic scattering cross section is folded with the (experimentally extracted) parton distribution function (PDF) inside the hadron and with the parton to hadron fragmentation function (FF) by using the factorization theorem in QCD \cite{fcw,fc2w} to calculate the hadron production cross section which is experimentally measured at the high energy colliders.

As mentioned above the parton distribution function (PDF) inside the hadron and the parton to hadron fragmentation function (FF) are experimentally extracted because the PDF and FF are not calculated yet. The hadron formation from the quarks, antiquarks and gluons is a long distance phenomenon in QCD where the renormalized QCD coupling becomes large. Since the pQCD is not applicable at the large distance, the PDF and FF cannot be calculated by using the pQCD.

At the large distance the non-perturbative QCD is necessary to study the parton distribution function (PDF) inside the hadron and the parton to hadron fragmentation function (FF). However, the analytical solution of the non-perturbative QCD is not known yet because of the presence of the cubic and quartic gluonic field terms in the QCD lagrangian inside the path integration in QCD (see section II for details). Due to this reason the PDF and FF are not calculated yet and hence they are extracted from the experiments.

However, the path integration in QCD can be performed numerically in the Euclidean time by using the lattice QCD method. Hence the lattice QCD studies the hadron formation from the quarks and gluons by evaluating the non-perturbative partonic correlation function in QCD numerically in the Euclidean time \cite{nkca}.
Since the lattice QCD can study the non-perurbative QCD one expects that the parton distribution function (PDF) inside the hadron and the parton to hadron fragmentation function (FF) can be calculated by using the lattice QCD method. Recently there has been lot of progress in the study of the parton distribution function (PDF) inside the hadron by using the lattice QCD method \cite{nklat}.

However, it is claimed in the literature that the parton to hadron fragmentation function (FF) cannot be studied by using the lattice QCD method because the fragmentation function involves the sum of outgoing unobserved (inclusive) hadrons.

But in this paper we find that since the hadron formation from the partons can be studied by using the lattice QCD method \cite{nkca,nkca1}, the parton to hadron fragmentation function can be studied by using the lattice QCD method by using the LSZ reduction formula for the partonic processes.

The paper is organized as follows. In section II we discuss the lattice QCD method to study the hadron formation from quarks and gluons. In section III we describe the LSZ reduction formula for the partonic processes in the lattice QCD method. In section IV we describe the lattice QCD method to study the parton to hadron fragmentation function. Section V contains conclusions.

\section{ Lattice QCD Method To Study Hadron Formation From Quarks and Gluons }

The partonic operator ${\cal O}_{{\eta}_c}(x)$ for $\eta_c$ formation in the lattice QCD method is given by
\bea
{\cal O}_{{\eta}_c}(x) =\Psi^\dagger_i(x) \Psi_i(x)
\label{copa}
\eea
where $\Psi_i(x)$ is the charm quark field with color index $i=1,2,3$. Using the lattice QCD method we find that \cite{nkca}
\bea
&&|<\eta_c(P)|c{\bar c}>|^2=|<\eta_c(P)|{\cal O}_{{\eta}_c}(0)|0>|^2\nonumber \\
&&=\left[\frac{<0|\sum_{\vec x}e^{i{\vec P}\cdot {\vec x}} {\cal O}_{{\eta}_c}({\vec x},\tau){\cal O}_{{\eta}_c}(0)|0>}{e^{ [\frac{<0|\sum_{{\vec x}'}~e^{i{\vec P}\cdot {\vec x}'}{\cal O}_{{\eta}_c}({\vec x}',\tau') [\int d\tau \int d^3x \sum_{q,{\bar q}, g} \partial_n T^{n0}_{\rm Partons}({\vec x},\tau)] {\cal O}_{{\eta}_c}(0)|0>}{<0|\sum_{{\vec x}'}e^{i{\vec P}\cdot {\vec x}'}{\cal O}_{{\eta}_c}({\vec x}',\tau') {\cal O}_{{\eta}_c}(0)|0>}]_{\tau'\rightarrow \infty}}}\right]_{\tau \rightarrow \infty} e^{\tau E_{\eta_c}(P)}\nonumber \\
\label{fraga}
\eea
where $\tau$ is Euclidean time, the ${\vec P}$ is the momentum of $\eta_c$, the $E_{\eta_c}(P)$ is the energy of $\eta_c$, the $|0>$ is the non-perturbative QCD vacuum state, the $\int dt$ is indefinite integration, $\int d^3x$ is definite integration and $T^{\mu \nu}(x)$ is energy-momentum tensor density of the parton in QCD.  

In eq. (\ref{fraga})
\bea
&& <0|{\cal O}_{{\eta}_c}(x'){\cal O}_{{\eta}_c}(0)|0>=\frac{1}{Z[0]} \int [d{\bar \psi}_1][d\psi_1][d{\bar \psi}_2][d\psi_2][d{\bar \psi}_3][d\psi_3] [d{\bar \Psi}][d\Psi][dA] \times {\cal O}_{{\eta}_c}(x'){\cal O}_{{\eta}_c}(0)\nonumber \\
&&\times {\rm det}[\frac{\delta K_f^a}{\delta \omega^b}] \times~{\rm exp}[i\int d^4x [-\frac{1}{4} F_{\sigma \mu}^h(x)F^{\sigma \mu b}(x) -\frac{1}{2\alpha} [K_f^a(x)]^2 \nonumber \\
&&+\sum_{f=1}^3 {\bar \psi}^i_f(x)[\delta^{ik}(i{\not \partial}-m_f)+gT^b_{ik}\aslash^b(x)]\psi^k_f(x)+{\bar \Psi}_i(x)[\delta^{ik}(i{\not \partial}-M)+gT^b_{ik}\aslash^b(x)]\Psi_k(x)]]
\label{npea}
\eea
and
\bea
&& Z[0]= \int [d{\bar \psi}_1][d\psi_1][d{\bar \psi}_2][d\psi_2][d{\bar \psi}_3][d\psi_3] [d{\bar \Psi}][d\Psi][dA] \nonumber \\
&&\times {\rm det}[\frac{\delta K_f^a}{\delta \omega^b}] \times~{\rm exp}[i\int d^4x [-\frac{1}{4} F_{\sigma \mu}^h(x)F^{\sigma \mu b}(x) -\frac{1}{2\alpha} [K_f^a(x)]^2 +\sum_{f=1}^3 {\bar \psi}^i_f(x)[\delta^{ik}(i{\not \partial}-m_f)\nonumber \\
&&+gT^b_{ik}\aslash^b(x)]\psi^k_f(x)+{\bar \Psi}_i(x)[\delta^{ik}(i{\not \partial}-M)+gT^b_{ik}\aslash^b(x)]\Psi_k(x)]]
\label{z0a}
\eea
with
\bea
F_{\sigma \mu}^b(x) =\partial_\sigma A_\mu^b(x) - \partial_\mu A_\sigma^b(x)+gf^{bds} A_\sigma^d(x) A_\mu^s(x).
\label{fna}
\eea

In the above equations the $\psi^i_f$ is the light quark field of flavor $f=1,2,3$ (= up, down, strange quarks), $A_\mu^a(x)$ is the gluon field, $K_f^a(x)$ is the gauge fixing term [in covariant gauge it is given by $K_f^b(x)=\partial^\lambda A_\lambda^b(x)$], the $m_f$ is the mass of the light quark of flavor $f$, the $M$ is the mass of the heavy quark, $\alpha$ is gauge fixing parameter and we do not have any ghost field because we directly work in the ghost determinant ${\rm det}[\frac{\delta K_f^a}{\delta \omega^b}]$.

The heavy quarkonium $\eta_c$ formation from the charm-anticharm $c{\bar c}$ pair can be studied from eq. (\ref{fraga}) by using the lattice QCD method.

\section{ LSZ Reduction Formula For The Partonic Processes In The Lattice QCD Method }

For simplicity we consider the $q \rightarrow \eta_c$ fragmentation function using lattice QCD method in this paper where $q$ is the light quark. Extension of this procedure to other parton to hadron fragmentation function is straightforward. 

Let us consider the parton to hadron fragmenting process
\bea
q \rightarrow \eta_c + X
\label{etca}
\eea
where $X$ is the unobserved particle(s). The partonic process corresponding to eq. (\ref{etca}) is given by
\bea
q(p_1)\rightarrow {\bar c}(p_2)c(p_3)+X.
\label{sca}
\eea
By using the LSZ reduction formula we find that the transition amplitude $<c{\bar c}+X|q>$ for the process in eq. (\ref{sca}) is given by \cite{nklsz}
\bea
&&<c{\bar c}+X|q> =(-i)i^2 \int d^4x_3 \int d^4x_2 \int d^4x_1 e^{ip_3 \cdot x_3+ip_2 \cdot x_2-ip_1 \cdot x_1} \times {\bar u}(p_3,\lambda_3)v(p_2,\lambda_2)u(p_1,\lambda_1)\nonumber \\ && \times \int d^4y_3 \int d^4y_2 \int d^4y_1 [G_R^c(x_3,y_3)]^{-1}[G_R^c(x_2,y_2)]^{-1}[G_R^q(x_1,y_1)]^{-1}<X|\Psi(y_3) {\bar \Psi}(y_2) \psi(y_1)|0>_R\nonumber \\
\label{lszac}
\eea
where $|0>$ is the non-perturbative QCD vacuum state, $\lambda$ represents the helicity, $\psi$ is the quark field of the light quark $q$, the $\Psi$ is the heavy (charm) quark field and $R$ stands for renormalized quantities. 

In eq. (\ref{lszac}) the $[G(x',x'')]^{-1}$ in coordinate space is defined by
\bea
\int d^4z' G(y''',z')[G(z',y'')]^{-1}=\delta^{(4)}(y'''-y'')
\label{grfg}
\eea
where
\bea
&& G^c(x',x'')
=\frac{1}{Z[0]} \int [d{\bar \psi}_1][d\psi_1][d{\bar \psi}_2][d\psi_2][d{\bar \psi}_3][d\psi_3] [d{\bar \Psi}][d\Psi][dA] \nonumber \\
&& \times {\bar \Psi}(x') \Psi(x'')\times {\rm det}[\frac{\delta K_f^a}{\delta \omega^b}] \times~{\rm exp}[i\int d^4x [-\frac{1}{4} F_{\sigma \mu}^h(x)F^{\sigma \mu b}(x) -\frac{1}{2\alpha} [K_f^a(x)]^2 +\sum_{f=1}^3 {\bar \psi}^i_f(x)\nonumber \\
&&[\delta^{ik}(i{\not \partial}-m_f)+gT^b_{ik}\aslash^b(x)]\psi^k_f(x)+{\bar \Psi}_i(x)[\delta^{ik}(i{\not \partial}-M)+gT^b_{ik}\aslash^b(x)]\Psi_k(x)]]
\label{npea2}
\eea
and
\bea
&& G^q(x',x'')
=\frac{1}{Z[0]} \int [d{\bar \psi}_1][d\psi_1][d{\bar \psi}_2][d\psi_2][d{\bar \psi}_3][d\psi_3] [d{\bar \Psi}][d\Psi][dA] \nonumber \\
&& \times {\bar \psi}(x') \psi(x'')\times {\rm det}[\frac{\delta K_f^a}{\delta \omega^b}] \times~{\rm exp}[i\int d^4x [-\frac{1}{4} F_{\sigma \mu}^h(x)F^{\sigma \mu b}(x) -\frac{1}{2\alpha} [K_f^a(x)]^2 +\sum_{f=1}^3 {\bar \psi}^i_f(x)\nonumber \\
&&[\delta^{ik}(i{\not \partial}-m_f)+gT^b_{ik}\aslash^b(x)]\psi^k_f(x)+{\bar \Psi}_i(x)[\delta^{ik}(i{\not \partial}-M)+gT^b_{ik}\aslash^b(x)]\Psi_k(x)]].
\label{npea3}
\eea

From eq. (\ref{lszac}) we find that the probability $P_{q\rightarrow {\bar c}c}$ for the light quark $q$ to fragment to $ {\bar c}c$ is given by
\bea
&& P_{q\rightarrow {\bar c}c}=N\sum_X <q|{\bar c}c,X><X,c{\bar c}|q>=N\int d^4x_3 \int d^4x_2 \int d^4x_1 \int d^4x'_3 \int d^4x'_2 \int d^4x'_1 \nonumber \\
&&e^{ip_3 \cdot (x_3-x_3')+ip_2 \cdot (x_2-x_2')-ip_1 \cdot (x_1-x_1')} \times [\sum_{\rm spin} |{\bar u}(p_3,\lambda_3)v(p_2,\lambda_2)u(p_1,\lambda_1)|^2]\times \int d^4y_3 \int d^4y_2 \int d^4y_1  \int d^4y'_3 \nonumber \\ && \int d^4y'_2 \int d^4y'_1 [G_R^c(x_3,y_3)]^{-1}[G_R^c(x_2,y_2)]^{-1}[G_R^q(x_1,y_1)]^{-1}[G_R^c(x'_3,y'_3)]^{-1}[G_R^c(x'_2,y'_2)]^{-1}[G_R^q(x'_1,y'_1)]^{-1}\nonumber \\
&& \sum_X <0|\psi(y'_1) {\bar \Psi}(y'_2) \Psi(y'_3)|X><X|\Psi(y_3) {\bar \Psi}(y_2) \psi(y_1)|0>_R
\label{lsz1}
\eea
which gives
\bea
&& P_{q\rightarrow {\bar c}c}=N\int d^4x_3 \int d^4x_2 \int d^4x_1 \int d^4x'_3 \int d^4x'_2 \int d^4x'_1 e^{ip_3 \cdot (x_3-x_3')+ip_2 \cdot (x_2-x_2')-ip_1 \cdot (x_1-x_1')}\nonumber \\
&&\times [ \sum_{\rm spin}|{\bar u}(p_3,\lambda_3)v(p_2,\lambda_2)u(p_1,\lambda_1)|^2] \int d^4y_3 \int d^4y_2 \int d^4y_1  \int d^4y'_3 \int d^4y'_2 \nonumber \\ && \int d^4y'_1 [G_R^c(x_3,y_3)]^{-1}[G_R^c(x_2,y_2)]^{-1}[G_R^q(x_1,y_1)]^{-1}[G_R^c(x'_3,y'_3)]^{-1}[G_R^c(x'_2,y'_2)]^{-1}[G_R^q(x'_1,y'_1)]^{-1}\nonumber \\
&& <0|\psi(y'_1) {\bar \Psi}(y'_2) \Psi(y'_3)\Psi(y_3) {\bar \Psi}(y_2) \psi(y_1)|0>_R
\label{lsz2}
\eea
where $N$ is the normalization factor given by eq. (\ref{nf}) and
\bea
&&<0| \psi(y'_1) {\bar \Psi}(y'_2) \Psi(y'_3)\Psi(y_3) {\bar \Psi}(y_2) \psi(y_1)|0>=\frac{1}{Z[0]} \int [d{\bar \psi}_1][d\psi_1][d{\bar \psi}_2][d\psi_2][d{\bar \psi}_3][d\psi_3] [d{\bar \Psi}][d\Psi][dA] \nonumber \\
&& \times \psi(y'_1) {\bar \Psi}(y'_2) \Psi(y'_3)\Psi(y_3) {\bar \Psi}(y_2) \psi(y_1)\times {\rm det}[\frac{\delta K_f^a}{\delta \omega^b}] \times~{\rm exp}[i\int d^4x [-\frac{1}{4} F_{\sigma \mu}^h(x)F^{\sigma \mu b}(x) -\frac{1}{2\alpha} [K_f^a(x)]^2 \nonumber \\
&&+\sum_{f=1}^3 {\bar \psi}^i_f(x)[\delta^{ik}(i{\not \partial}-m_f)+gT^b_{ik}\aslash^b(x)]\psi^k_f(x)+{\bar \Psi}_i(x)[\delta^{ik}(i{\not \partial}-M)+gT^b_{ik}\aslash^b(x)]\Psi_k(x)]].
\label{npea1}
\eea
Note that in eq. (\ref{lsz2}) the normalization factor $N$ arises because the creation and annihilation operators in the derivation of the LSZ reduction formula are not dimensionless \cite{psa}. Due to this reason $\sum_X <q|{\bar c}c+X><X,c{\bar c}|q>=\sum_X |<X+c{\bar c}|q>|^2$ is not dimensionless. Hence one finds that the normalization factor $N$ is given by
\bea
N=\frac{1}{<q|q><{\bar c}c|c{\bar c}>}
\label{nf}
\eea
where
\bea
&&<q|q> = \int d^4x'_1 \int d^4x_1 e^{ip_1 \cdot (x'_1-x_1)} \times [\sum_{\rm spin}{\bar u}(p_1,\lambda_1)u(p_1,\lambda_1)]\times \int d^4y'_1 \nonumber \\ && \int d^4y_1 [G_R^q(x'_1,y'_1)]^{-1}[G_R^q(x_1,y_1)]^{-1}<0|{\bar \psi}(y'_1) \psi(y_1)|0>_R
\label{qnf}
\eea
and
\bea
&&<c{\bar c}|c{\bar c}> = \int d^4x_3 \int d^4x_2 \int d^4x'_3 \int d^4x'_2 e^{ip_3 \cdot (x_3-x'_3)+ip_2 \cdot (x_2-x'_2)} \times [\sum_{\rm spin} {\bar u}(p_3,\lambda_3) u(p_3,\lambda_3)]\nonumber \\
&&[\sum_{\rm spin} v(p_2,\lambda_2) {\bar v}(p_2,\lambda_2)]  \int d^4y_3 \int d^4y_2 \int d^4y'_3 \int d^4y'_2 [G_R^c(x_3,y_3)]^{-1}[G_R^c(x_2,y_2)]^{-1}[G_R^c(x'_3,y'_3)]^{-1}\nonumber \\
&&[G_R^c(x'_2,y'_2)]^{-1}<0|\Psi(y_3) {\bar \Psi}(y_2) \Psi(y'_2) {\bar \Psi}(y'_3)|0>_R.
\label{cnf}
\eea

Since the path integrations in eqs. (\ref{npea2}), (\ref{npea3}), (\ref{qnf}), (\ref{cnf}) and (\ref{npea1}) can be performed numerically by using the lattice QCD method in the Euclidean time to calculate the non-perturbative correlation functions $G^c(x,x')$, $G^q(x,x')$, $<0|{\bar \psi}(y'_1) \psi(y_1)|0>$, $<0|\Psi(y_3) {\bar \Psi}(y_2) \Psi(y'_2) {\bar \Psi}(y'_3)|0>$ and $<0| \psi(y'_1) {\bar \Psi}(y'_2) \Psi(y'_3)\Psi(y_3) {\bar \Psi}(y_2) \psi(y_1)|0>$ respectively one finds from eq. (\ref{lsz2}) that the probability $P_{q\rightarrow {\bar c}c}$ of the light quark $q$ to fragment to $c{\bar c}$ can be calculated by using the lattice QCD method.

\section{ Lattice QCD Method To Study Parton To Hadron Fragmentation Function }

From eqs. (\ref{fraga}) and (\ref{lsz2}) we find that the $q \rightarrow \eta_c$ fragmentation function $D_{q\rightarrow \eta_c}(P)$ in the lattice QCD method is given by
\bea
&&D_{q\rightarrow \eta_c}(P)= P_{q\rightarrow {\bar c}c}\times |<\eta_c(P)|c{\bar c}>|^2
\label{fragf}
\eea
where the probability $P_{q\rightarrow {\bar c}c}$ is given by eq. (\ref{lsz2}) and $|<\eta_c(P)|c{\bar c}>|^2$ is given by eq. (\ref{fraga}).

Since $P_{q\rightarrow {\bar c}c}$ in eq. (\ref{lsz2}) can be calculated by using the lattice QCD method in the Euclidean time and the $|<\eta_c(P)|c{\bar c}>|^2$ can be calculated from eq. (\ref{fraga}) by using the lattice QCD method in the Euclidean time one finds from eq. (\ref{fragf}) that the fragmentation function $D_{q\rightarrow \eta_c}(P)$ can be calculated by using the lattice QCD method in the Euclidean time. 

Extension of eq. (\ref{fragf}) to other parton to hadron fragmentation function is straightforward. 

\section{Conclusions}
In the literature it is assumed that the parton to hadron fragmentation function cannot be studied by using the lattice QCD method because of the sum over the (unobserved) outgoing hadronic states. However, in this paper we have found that since the hadron formation from the partons can be studied by using the lattice QCD method, the parton to hadron fragmentation function can be studied by using the lattice QCD method by using the LSZ reduction formula for the partonic processes.

\end{document}